\documentclass[times, 10pt, twocolumn]{article}
\usepackage{latex8}
\usepackage{times}
\usepackage{algorithm}
\usepackage{algorithmicx}
\usepackage{cite}
\usepackage{algpseudocode}
\usepackage{hyperref}
\usepackage{multirow}
\usepackage{graphicx}
\usepackage{listings}
\usepackage{booktabs}

\begin{document}
%
\title{Testing as an Investment}

\author{Xiaoran Xu$^{123}$, Chunrong Fang$^{12}$, Zhenyu Chen$^{12}$*\\
$^1$State Key Laboratory for Novel Software Technology, Nanjing University, Nanjing, China\\
$^2$Software Institute, Nanjing University, Nanjing, China\\
$^3$Computer Science Department, Rice University, Houston, Texas, USA\\
$^*$Corresponding author: pyzychen@gmail.com\\
}


\maketitle

\begin{abstract}
Software testing is an expensive and important task. Plenty of researches and industrial efforts have been invested on improving software testing techniques, including criteria, tools, etc. These studies can provide guidelines to select suitable test techniques for software engineers. However, in some engineering projects, business issues may be more important than technical ones, hence we need to lobby non-technical members to support our decisions. In this paper, a well-known investment model, Nelson-Siegel model, is introduced to evaluate and forecast the processes of testing with different testing criteria.  Through this model, we provide a new perspective to understand short-term, medium-term, and long-term returns of investments throughout the process of testing. A preliminary experiment is conducted to investigate three testing criteria from the viewpoint of investments. The results show that statement-coverage criterion performs best in gaining long-term yields; the short-term and medium-term yields of testing depend on the scale of programs and the number of faults they contain.
\end{abstract}

\textbf{Keywords: Investment, Nelson-Siegel model, Cost-yields, Testing criterion.}

%
\section{Introduction}
Testing is an important and labor-intensive activity in software development life cycle. It always consumes from 30\% to 50\% of total development cost according to [1]. Many studies have been conducted to compare various testing criteria and tools aiming at providing guidelines of testing for practitioners[15]-[18]. Furthermore, a number of models, such as APFD, were proposed to evaluate testing criteria [2]. Commonly, these studies concern about the technical issues, such as fault detection capability, execution time, etc. However, in industries, it is the customers that invest and drive software development. For these customers, business issues are more significant than technical ones. Therefore, we need to provide business evidences rather than technical guidelines for non-technical members in companies.

Investment knowledge is a branch of economics. It concerns about how to invest limited resources of individuals or organizations to financial assets, like stock, national debt and real estate, in order to gain reasonable cash flow and risk-benefit ratio. The key point of investment knowledge is finding the optimal equilibrium solution of personal assets allocation, under the guidelines of utility maximization criterion. Investment philosophy is widely accepted in daily life, among which Return On Investment (ROI) [3] is a way to consider profits of capital investment. Many ROI metrics were proposed in the past century[19], which are widely used to support marketing decisions in a lot of industries, including software industry [4]. In this paper, we innovatively propose the concept that testing is an investment and introduce a well-known model, Nelson-Siegel (NS) model [5], to calculate ROI of several testing criteria. NS model is a classical model to estimate the current term structure of interest and forecast the future term structure, which has been applied in many fields of finance [6][7]. In this paper, we will use it to model the relationship between investments and returns of testing.

The main contributions of this paper are as following:
\begin{itemize}
\item To the best of our knowledge, it is the first time to use investment model to quantitatively investigate investments and returns of software testing.
\item An experiment is conducted to analyze the yields of test activity and compare the yields among testing criteria: random, statement-coverage and branch-coverage.
\end{itemize}

The application of NS model has shown a potential prospect albeit our experiments just achieve a fundamental stage. The NS model, equipped with total economic indices such as fault price and test cost, could completely reflect software testing to investment area. Thus a novel, practical and comprehensive testing evaluation method is created. This model can firstly evaluate and forecast the yields of software testing, secondly explore the greatest yields point and moreover calculate the cost and risk of remained faults according to the maximum long-term yields prediction. Details will be discussed in Section 4.

The rest of this paper is organized as follows. Section 2 proposes the detail of our approach. The experiment design and the result analysis are presented in Section 3. We discuss the applicability of NS model in testing and some related parameters in Section 4.  The conclusion and future work are presented in Section 5 and Section 6, respectively.

\section{Approach}
The NS model is proposed to optimally estimate, model and forecast the term structure of interest rates using a flexible, smooth and parametric function. It is capable of capturing many of the typically observed shapes that the yield curve assumes and is usually used in investment subject. If the instantaneous forward rate at maturity $m$, denoted $r(m)$, is given as 
\begin{equation}
\label{eq:eq1}
r(m) = \beta_0 + \beta_1 * e^{-m/\tau} +  \beta_2 [\frac{m}{\tau}*e^{-m/\tau}]~.
\end{equation}
~\\
Then the yield to maturity, denoted $R(m)$, is the average of the forward rates
\begin{equation}
\label{eq:eq2}
R(m) = \frac{1}{m}\int_{0}^{m}r(x)dx~.
\end{equation}
~\\
So in Eq.1 the resulting function of integrating $r(m)$ from zero to $m$ and being divided by $m$ (Eq.2) is
\begin{equation}
\label{eq:eq3}
R(m) = \beta_0 + \beta_1 \frac{1-e^{-\frac{m}{\tau}}}{\frac{m}{\tau}} + \beta_2 \left[\frac{1-e^{-\frac{m}{\tau}}}{\frac{m}{\tau}} - e^{-\frac{m}{\tau}}\right]~,
\end{equation}
~\\
among which $\beta_0$, $\beta_1$, $\beta_2$ and $\tau$ are parameters to be estimated. It is easy to prove that 
\begin{equation}
\label{eq:eq4}
r(+\infty) = \beta_0~,
\end{equation}
~\\
and 
\begin{equation}
\label{eq:eq5}
r(+\infty) - r(0) = - \beta_1~.
\end{equation}
~\\
Thus, $\beta_0$, $\beta_1$, $\beta_2$ are respectively the Level Factor, Slope Factor (the absolute value of $\beta_1$) and Curvature Factor of the yield curve. From the perspective of dynamic period yields, they are also long-term, short-term, and medium-term components. The parameter $\tau$ controls the speed of other parameters�� decaying, especially the extreme value of the loading to which $\beta_2$ attaches.

Assuming it can be applied to testing, NS model will be able to model, evaluate and forecast the structure of testing yield curve. To simplify and avoid contingent price difference, we assume that every test case in our approach has the same investment-one unit and, similarly, every fault has one unit return. Thus we respectively transform test cases and faults to testing investments and returns. In NS model, $m$ refers to testing investments and $R(m)$ stands for testing returns. Furthermore, $\beta_0$ which is a constant, can be regarded as the maximum expected long-term yields in testing activity. $\beta_1$ and $\beta_2$ respectively  relate to the short-term and medium-term yields of testing, namely, the faster the curve reaches the stable limit, the faster the returns are gained; the more convex the curve is, the greater the returns are. Since $\tau$ controls the extreme value point of the loading to which $\beta_2$ attaches and $\beta_2$ has a completely positive correlation with the curvature of NS model curve, $\tau$ refers to the point with extreme yields, which can be treated as the primary testing stopping point.

Nevertheless, the basic problem is whether the NS model can be applied to testing activities as we assumed. Our experiments of curve fitting described in Section 3, used nonlinear least square method on two programs with more than 51,000 lines of codes, and showed that approximately 83\% percent of the values of correlation coefficient $R^2$ are beyond 0.9. This result verifies our assumption that NS model can be used in testing.

\section{Experiment}

\subsection{Subject Programs}
Two programs are used in our empirical study: NanoXML (version 2)\footnote{http://sir.unl.edu/portal/index.php} and Checkstyle (version 5.3)\footnote{http://checkstyle.sourceforge.net}. NanoXML is a small XML parser for Java and it is an easy-to-use, non-GUI based system, which can be built from source without external libraries. Checkstyle is a development tool to help programmers write Java code that adheres to a coding standard. The version of each program is selected randomly. These two programs are both real world open source programs and have self-validating test cases available. The details of subject programs are listed in Table 1.

\newsavebox{\tablebox}
\begin{table}[!t]
\caption{Subject Programs}
\center
\label{tab:subject}
\begin{lrbox}{\tablebox}
\begin{tabular}{|c|c|c|c|c|}
\hline
Program	&	NanoXML	& Checkstyle \\
\hline
LOC & 7646	& 43407 \\
\hline
\# Branches &	26 &	2449 \\
\hline
\# Faults &	7 &	687 \\
\hline
\# Test Cases &	214 &	162 \\
\hline
\end{tabular}
\end{lrbox}
\scalebox{1}{\usebox{\tablebox}}
\end{table}

\subsection{Test Cases}
Test cases are downloaded together with corresponding programs and they are all JUnit test cases. One test case has been split into several smaller ones according to every independent testing method and the effectiveness of original test suite will not change since no test method has been added or removed. 

In addition, the tool CodeCover\footnote{http://http://codecover.org} has been equipped on the source codes to help collect coverage information of a System Under Test (SUT). CodeCover is an open source testing tool which can measure statement, branch and MC/DC coverage. The information of branches and conditions can be accessed by the intermediate files generated by CodeCover. Thus we can generate coverage vector of each test case: respectively, use 0 and 1 to represent certain statement or branch is covered or not. 

Then Additional Greedy Algorithm is utilized to pick test cases satisfied statement-coverage or branch-coverage criteria. The fundamental logic of Additional Greedy Algorithm is to select the next test case that covers the maximum number of statements or branches that have not yet been covered in past selections. The reason why we choose Additional Greedy Algorithm is that it has been proven of high effectiveness in a series of coverage-based test case prioritization techniques [13]. In addition, some other greedy algorithms, such as 2-Optimal Algorithm, Hill Climbing, Genetic Algorithms and so forth, cannot outperform significantly than the Additional Greedy Algorithm, which is considered as the cheaper-to-implement-and- execute algorithm [14]. Therefore, we chose to implement this algorithm.

\subsection{Faults}
Faults in NanoXML are all original faults while in Checkstyle are mutations. In [8], Andrews et al. verifies the feasibility of using mutation instead of real faults. Mutations of Checkstyle are generated by the tool Jumble\footnote{http://jumble.sourceforge.net/}. Jumble is a class level mutation testing tool which is integrated with JUnit. The number of mutants depends on the number of test cases but not the size of programs since Jumble is associated with JUnit. Testing reports are produced by Jumble automatically so that we can extract the data and then set up fault vectors: using 0 and 1 to refer that this fault has been detected or not, the same principle with coverage vectors.

\subsection{Curve Fitting of Nelson-Siegel Model}
The numbers of test cases which are qualified by random, branch-coverage or statement-coverage criterion are counted and recorded as the variable $m'$. Then $m'$ is converted to the independent variable $m$ followed the principle that one test case merits one unit in investment. So does the corresponding number of faults-detected to dependent variable $R(m)$. The nonlinear least square method is performed to fit the NS model curve, then the value of correlation coefficient $R^2$ and parameters to be estimated, namely $\beta_0$, $\beta_1$ ,$\beta_2$ and $\tau$, are recorded.

\subsection{Experiment Setup}
The main process of experiments we design and conduct is shown in Fig.1, which contains the following steps: (1) Instrument subject programs using CodeCover; (2) Run test cases of subject programs against instrumented version; (3) Collect coverage information to generate coverage vectors, including the granularities of statement and branch; (4) Run test cases of subject programs; (5) Collect mutation information and generate fault vectors; (6) Pick test cases which satisfy random, branch-coverage or statement-coverage criteria through Additional Greedy Algorithm; (7) Convert data from step 5 and 6 to testing investments and returns; (8) Operate curve fitting on NS model using the statistic data from step 7.

\begin{figure}[!t]
\centering
\scalebox{0.4}[0.4]{
\includegraphics{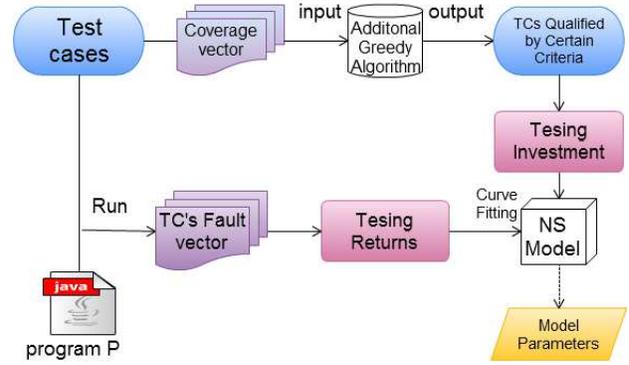}
}
\caption{Framework of the Experiment}
\label{fig:coupling}
\end{figure}

\subsection{Results}
The values of parameters $\beta_0$, $\beta_2$, $\tau$, the the absolute value of $\beta_1$ and the correlation coefficient $R^2$ are listed in Table 2. R, S and B respectively represent random, statement-coverage and branch-coverage criterion.

\begin{table}[!t]
\centering
\caption{Vaules of Parameters in NS Model and Correlation Coefficient $R^2$}
\center
\label{tab:results}
\begin{lrbox}{\tablebox}
\begin{tabular}{|c|c|c|c|c|c|c|}
\hline
Program      & Criterion & $R^2$ & $\beta_0$	&  $|\beta_1|$& $\beta_2$ & $\tau$ \\
\hline
           	   & R & 0.988	& 411.40 &	435.3 & 759.5	& 69.57\\
\cline{2-7}
Checkstyle & B & 0.991	& 594.20 &	607.9 & 339.2   & 55.81\\
\cline{2-7}
		  & S	& 0.977     &613.80  & 616.9 & -429.2	& 19.82\\
\hline
		  &R	& 0.846	& 6.86 &	8936.0  & 8902.0	& 1.01\\
\cline{2-7}
NanoXML   & B & 0.959	& 7.21 &7.4  & -8.6  & 1.66\\
\cline{2-7}
	           & S & 0.941	& 7.39 &	6.3   & -13.1  & 14.95\\
\hline
\end{tabular}
\end{lrbox}
\scalebox{0.8}{\usebox{\tablebox}}
\end{table}

\begin{figure}[!t]
\centering
\scalebox{0.5}[0.5]{
\includegraphics{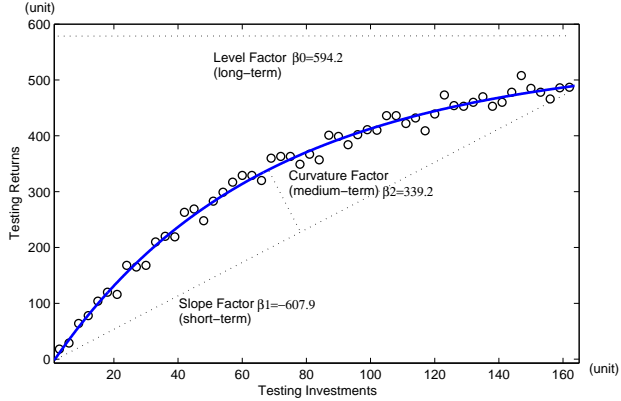}
}

\caption{Values of Parameters in Checkstyle, Branch-Coverage Criterion}
\label{fig:apfdall}
\end{figure}

\begin{figure}[!t]
\centering
\scalebox{0.5}[0.5]{
\includegraphics{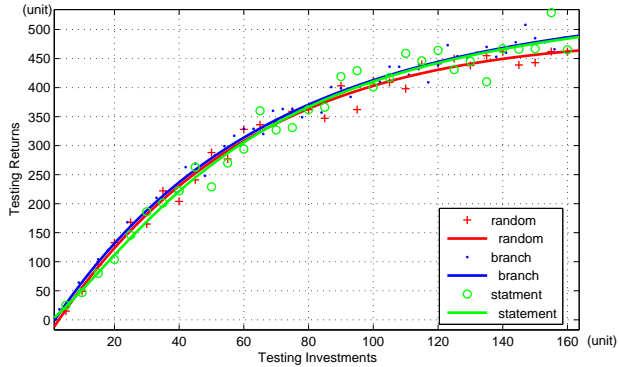}
}

\caption{Curve Fitting Result of Checkstyle}
\label{fig:apfdall}
\end{figure}

In Table 2, it is observed that 5 of 6 correlation coefficient $R^2$ are greater than 0.94 and the rest is 0.846. In Fig. 2, the detail of the curve fitting on Checkstyle is depicted as an example. And Fig. 3 marks the parameters of curve fitting on Checkstyle branch-coverage criterion testing. For the limited space, only Checkstyle is shown.

The maximum expected long-term yields component $\beta_0$ tends to be the smallest in random testing and the largest in statement-coverage criterion testing for both programs. For Checkstyle, the absolute value of short-term component $\beta_1$ has a regulation that S$>$B$>$R and the parameter $\tau$, related to the extreme yields point, appears to be S$<$B$<$R. Interestingly, in NanoXML, $\beta_1$ and $\tau$ present totally reversed trends. At last, the medium-term yields component $\beta_2$ are positive in R, negative in S for both programs and has different sign in B.

\section{Discussion}
\subsection{Applicability of NS Model in Testing}
From the experiment results in Section 3 that 5 of 6 correlation coefficient $R^2$ are greater than 0.94 and the rest is 0.846, it can be safely concluded that the NS model is able to fit extremely well on the relationship between testing investments and returns. In other words, our introduction of NS model to testing is feasible and all of the following respective discussions of four parameters have reasonable and reliable fundament.

\subsection{The Parameter $\beta_0$}
As mentioned in Eq.4, the parameter $\beta_0$ is the stable limit of $R(m)$ and it means the maximum expected long-term yields in software testing. The experiment results show that, when investments are saturated, statement-coverage outperforms branch-coverage criterion and the latter outperforms random criterion in long-term yields. Comparatively, in most of previous researches, branch-coverage criterion tends to detect more faults than statement-coverage criterion since it has fine-grained requirements on coverage [9] [10]. However, stricter requirements naturally mean more test cases which lead to more investments. Compared with branch-coverage (high investments and high returns) and random (low investments and low returns), statement-coverage criterion which acquires a balance turns out to be the greatest yields choice.

\subsection{The Parameter $\tau$}
As explained in Section 2, the parameter $\tau$  refers to the maximum yields point which can be treated as the primary testing stopping point. In Checkstyle, the value of $\tau$ is S$<$B$<$R, which means the statement Ccoverage criterion will reach the extreme yields point at first, followed with branch and random. However, it is interesting that the order of the value of $\tau$ is just reversed in NanoXML. Considering the difference of program scale and the number of faults between Checkstyle and NanoXML, the reason might be that there are just 7 faults in NanoXML such that random test cases can detect them easily. At the same time, NanoXML has 7,646 lines of codes which means plenty of test cases are required to satisfy statement-coverage requirement, among which a large percentage is non-fault-detected. So the yields of statement-coverage criterion in NanoXML come later. On the contrary, Checkstyle has as many faults as 687 and its ratio of LOC to the number of faults is pretty smaller than the one in NanoXML. In other words, faults in Checkstyle are much more saturated than faults in NanoXML. So the advantage of random criterion that innately has the even-distributed probability in covering each fault, fades away and more coverage of statements stand out. To sum up, when testing investment is preliminary, it is recommended to use random criterion in fault-unsaturated programs while statement-coverage criterion in relatively fault-saturated programs, according to the parameter $\tau$.

\subsection{The Parameter $\beta_1$}
The prophase yields of testing is positively related to the primary testing stopping point. That means the short-term component $\beta_1$ should present consistent meaning with $\tau$ and our experiment results just verified it.  In terms of the absolute value of $\beta_1$, the bigger it is, the greater the short-term yields are. So in Checkstyle, the absolute value of $\beta_1$ is S$>$B$>$R, which means the prophase yields is S$>$B$>$R. It is indeed consistent with the appearing order of the primary testing stopping points. And this consistency is also suitable in NanoXML.

\subsection{The Parameter $\beta_2$}
The medium-term loading that $\beta_2$ attaches is a function that starts out at zero, increases at medium maturities and decays to zero. Therefore, $\beta_2$  is designed to create a hump-shape. Similar to the meaning of $\beta_1$, the medium-term component $\beta_2$ has a positive correlation with the medium-term yields of testing. In our experiment, the value of $\beta_2$ is positive in R, negative in S for both programs and has different signs in B. As we all know, random criterion is unconcerned with the structure of programs, so it displays a stable medium-term yields. In addition, it is supposed that the fluctuation in medium-term yields causes the negative value of $\beta_2$. Studies about $\beta_2$ could be paid further efforts.

Another interesting observation in our experiment results is that in all of the yields comparison according to four parameters, branch-coverage criterion always lies in the intermediate position, namely it performs neither the best nor the worst.
Therefore, if certain testing situation needs a trade-off among short-term, medium-term, and long-term yields, branch-coverage criterion will be a preferable choice.

As a preliminary experiment, only two subject programs are involved in our study, which can be extended in the future. However, these subject programs, which are of reasonable scale, are widely used in other studies like [11][12], and are representative to draw meaningful conclusions.

\section{Conclusion}
Testing is as an investment. Focusing on the business issues, investments and returns, rather than technical issues of software testing, we firstly introduce the model Nelson-Siegel to explore a novel and practical perspective of testing. Then, an experiment is conducted on two programs, with more than 51,000 lines of codes and nearly 700 faults, and the results indicate that the NS model can fit well on the relationship between investments and returns. Some useful observations are summarized as follows.
\begin{itemize}
\item Statement-coverage criterion has stronger ability to earn long-term yields than branch-coverage and random criterion.
\item For large-scale programs with saturated faults, statement-coverage criterion tends to reach primary testing stopping point at first and has the best prophase yields, while random criterion does the best in small-scale programs with unsaturated faults.
\item Random criterion outperforms the other two criteria in the medium-term yields of testing.
\item Branch-coverage criterion can be a trade-off in choice of combining all of the short-term, medium-term, and long-term yields.
\end{itemize}

\section{Future Work}
Our experiment is preliminary but encouraging. The prospect of testing as an investment and the application of the NS model are promising, furthermore, there are some aspects that could be improved and studied in the future.
\begin{itemize}
\item Some parameters of the NS model should be confined within a concrete range to exclude non-practical situation such as negative investments.
\item The transformation from testing data to economic data should be studied. To be more practical, the cost should take generation, execution and inspection of testing into calculation. Faults should be classified based on significance and type like GUI fault or logic fault. Practitioners who use this NS model can adjust the indices according to given programs.
\item A base line can be depicted which means no gain and no lose to be compared with certain testing curve to figure out profit or deficit.
\item NS model can be used to predict the risk in a released version. The number of expected-detected faults derived from $\beta_0$ subtracts the number of have-already-detected faults equals the number of have-not-detected faults, which remain in codes.  In other words, we have already got current yields and could predict the stable long-term yields, so the gap between them is the risk remained in a released version.
\end{itemize}

\section*{Acknowledgment}
The work described in this article was partially supported by the National Natural Science Foundation of China (61003024, 61170067) and the Scientific Research Foundation of Graduate School of Nanjing University (2013CL13).

\end{document}